\documentclass[12pt]{article}
\usepackage{epsfig}

\newcommand{\jpsi}{\ensuremath{J\!/\!\psi}}
\newcommand{\as}{\ensuremath{\alpha_s}}
\newcommand{\beq}{\begin{equation}}
\newcommand{\eeq}{\end{equation}}
\newcommand{\beqa}{\begin{eqnarray}}
\newcommand{\eeqa}{\end{eqnarray}}

\catcode`\@=11

\def\@citex[#1]#2{\if@filesw\immediate\write\@auxout{\string\citation{#2}}\fi
  \def\@citea{}\@cite{\@for\@citeb:=#2\do
    {\@citea\def\@citea{,\penalty\@m}\@ifundefined
       {b@\@citeb}{{\bf ?}\@warning
       {Citation `\@citeb' on page \thepage \space undefined}}%
\hbox{\csname b@\@citeb\endcsname}}}{#1}}

\def\citer{\@ifnextchar [{\@tempswatrue\@citexr}{\@tempswafalse\@citexr[]}}
\def\@citexr[#1]#2{\if@filesw\immediate\write\@auxout{\string\citation{#2}}\fi
  \def\@citea{}\@cite{\@for\@citeb:=#2\do
    {\@citea\def\@citea{--\penalty\@m}\@ifundefined
       {b@\@citeb}{{\bf ?}\@warning
       {Citation `\@citeb' on page \thepage \space undefined}}%
\hbox{\csname b@\@citeb\endcsname}}}{#1}}

\begin{document}


\begin{titlepage}

\vspace*{-5mm}

\begin{flushright}
DESY 96-147\\
RAL-96-094\\
November 1996
\end{flushright}

\vskip 1.25cm

\begin{center}
\boldmath
{\Large \bf Associated $\jpsi$ + $\gamma$ Photoproduction as a}\\[1mm] 
{\Large \bf Probe of the Colour-Octet Mechanism}
\unboldmath
\vglue 1.0cm
\begin{sc}
{\large\sc Matteo Cacciari$^a$, 
Mario Greco$^b$\
and\\[5pt] Michael Kr\"{a}mer$^c$}
\vglue 0.3cm
\end{sc}
$^a${\sl Deutsches Elektronen-Synchrotron DESY\\
D-22603 Hamburg, Germany}\\
$^b${\sl INFN, Laboratori Nazionali di Frascati and \\
Dipartimento di Fisica, Universit\`a di Roma III, Italy}\\
$^c${\sl Rutherford Appleton Laboratory\\
Chilton, Didcot, OX11 0QX, England}
\end{center}
\vglue 1.25cm

\begin{abstract} The associated production of a $\jpsi$ and a photon in
  photon-hadron collisions is considered, and shown to be a good probe
  of the presence of colour-octet mediated channels in quarkonia
  production, as predicted by the factorization approach within NRQCD.
  Total and differential cross sections for photoproduction at fixed
  target experiments and at HERA are presented.  Associated $\jpsi +
  \gamma$ production at hadron colliders is briefly discussed.
\end{abstract}

\vspace*{5mm}

\vfill
\noindent\rule{7cm}{.2mm}
\small
\vspace{-.3cm}
\begin{tabbing}
e-mail addresses: \= cacciari@desy.de \\
                  \> greco@lnf.infn.it \\
                  \> Michael.Kraemer@rl.ac.uk \\
\end{tabbing}

\end{titlepage}

\section{Introduction}
\label{intro}

The study of heavy quarkonia production is a good testing ground for
our understanding of the transition region between the realms of
perturbative and non-perturbative Quantum Chromodynamics (QCD).
Different production models have been devised, and all of them have to
deal with the problem of properly describing and matching these two
phases of QCD.  Most recently the so-called Non Relativistic QCD
(NRQCD) factorization approach (FA) has been proposed by Bodwin,
Braaten and Lepage (BBL) \cite{bbl}: an effective-field-theory
framework in NRQCD is used to separate the short-distance scale of
annihilation and production of heavy quarkonium from the
long-distance scales associated to the quarkonium structure.  A
brief review of the various models is presented in section
\ref{secpro}.

In particular the BBL approach (also referred to as FA) has been
applied to describe quarkonia production at the Tevatron \cite{cdf}
providing a successful picture \cite{cdfoctet} of the production rates
of $\jpsi$ and $\psi'$ - much larger than earlier predictions - which
is related to the relevance of colour-octet contributions in the
production mechanism. It is to be noted that this successful
description relies on a certain number of non-perturbative parameters
having to be fitted to the data. Therefore additional and more
extensive comparisons with experimental data coming also from other
kind of reactions are necessary to finally assess the validity of this
approach.  To this aim, calculations of inclusive quarkonia production
in $e^+e^-$ annihilation \cite{ee}, $Z$ decays \cite{zdecay}, hadronic
collisions at fixed-target experiments \cite{coft}, $B$ decays
\cite{bdecays} and $\gamma p$ collisions (both for $S$-wave
\citer{ck,gks} and $P$-wave \cite{ma,ckhera} states) have recently
been performed within its frame.  From refs.~\cite{ck,kls}, in
particular, the predictions based on the leading colour-octet
contribution appear not to agree with recent experimental data
obtained at HERA by the H1 Collaboration \cite{h1}, indicating either
a reduced phenomenological importance of the colour-octet terms than
suggested by the Tevatron data analysis \cite{cdfoctet} or the
possible relevance of higher order corrections.

In this paper we will consider the associated production of a $\jpsi$
and a photon, where the photon has to be directly produced in the
short-distance interaction, i.e. it should not come, for instance,
from the radiative decay of a $P$-wave quarkonium into a $\jpsi$.
This process therefore must have a very distinctive experimental
signature, namely a $\jpsi$ balanced in $p_T$ by the hard photon.  We
will argue that the study of this reaction in $\gamma p$ collisions
provides a powerful tool for assessing the importance of colour-octet
terms in quarkonia production. First results of our analysis have
already been presented in \cite{ckhera}.

The paper is organized as follows: in section \ref{secpro} we will
briefly review various quarkonia production models and the BBL
approach.  Then in section \ref{psigamma} we will introduce the basic
diagrams which contribute to $\jpsi + \gamma$ photoproduction and
discuss why this process provides a clear signature of colour-octet
terms. Numerical results and Conclusions will follow in sects. 4 and 5
respectively.

\section{Production Mechanisms}
\label{secpro}

Any approach attempting to describe the production (or decay) of a
heavy quarkonium must deal with two issues which - for sufficiently
large quark masses - can be kept distinct: the production (or the
annihilation) of the heavy quark-antiquark pair ($Q\bar Q$)
constituting the quarkonium and their binding into a single physical
long-lived particle.  The heavy quark mass, $m_Q \gg
\Lambda_{\mbox{\scriptsize QCD}}$, sets the scale for the
short-distance interaction of producing or annihilating a $Q\bar{Q}$
pair, which can therefore be studied within perturbative QCD (pQCD)
\cite{CSS}.  The binding of the two quarks into a bound state is on
the other hand a long-distance process, and non-perturbative effects
have to be included at this point. These are usually parametrized via
form factors describing the probability of the $Q\bar Q$ pair to form
the bound state. The degree of rigour, completeness and complexity of
this sector varies greatly from approach to approach:

The Colour Evaporation Model (CEM) \cite{CEM} rests on duality
arguments \cite{aeg} in assuming that $c\bar c$ pairs produced with an
invariant mass below that of a $D\bar D$ mesons threshold, (taking
charm as an example of heavy quark) will eventually hadronize into a
charmonium state. Being this assumption only qualitative this model of
course is unable to predict the production rate of a particular
quarkonium state. It is therefore not very suitable for the study of
exclusive final states.

The Colour Singlet Model (CSM) \cite{CSM} tries to overcome this
difficulty of the CEM by making a very precise request: the $Q\bar Q$
pair must be produced in the short-distance interaction with the same
spin, angular momentum and colour quantum numbers of the quarkonium
state.  A single parameter, provided by the bound state Bethe-Salpeter
wave function or its derivative, then describes the hadronization into
the observable particle. This model is of course much more predictive:
the production rate for a colour-singlet $Q\bar Q$ state with a given
spin and angular momentum, usually referred to as $Q\bar
Q[^{2S+1}L_J,\underline{1}]$, can be calculated exactly in pQCD.
Moreover, the phenomenological parameter can be measured, for
instance, in electromagnetic decays and used to make absolute
predictions for production rates. Despite its physical transparency
and predictive power, the CSM is clearly not a complete theory: It
ignores, at least in its original formulation, relativistic
corrections which take into account the nonzero relative velocity of
quark and antiquark in the bound state. Moreover, there are no
theorems that guarantee that the simple factorization of the
quarkonium production cross section into a short-distance part and a
single non-perturbative parameter is valid in higher orders of
perturbation theory. In fact, in the case of hadronic production or
decay of $P$-wave quarkonia, the radiative corrections to the
short-distance cross section contain infrared divergences that cannot
be factored into a single non-perturbative quantity. This failure is
to be attributed to an incomplete description of the quarkonium wave
function, in that the Colour Singlet Model ignores the contributions
from higher Fock-state components. As a result, CSM predictions for
producing $\jpsi$ and $\psi'$ states at large $p_T$ have been found to
grossly underestimate, by more than an order of magnitude, the
experimental cross sections obtained by the CDF
collaboration\cite{cdf} in $p\bar p$ collisions at the Tevatron (see
also \cite{mlm} and references therein).

The factorization approach by Bodwin, Braaten and Lepage \cite{bbl}
provides a rigorous framework for treating quarkonium production and
decays which resolves these problems. It extends the CSM by allowing
$Q\bar Q$ pairs with spin, angular momentum and colour quantum numbers
different from those of the observed quarkonium to hadronize into the
latter. In this respect, the factorization formalism recovers some of
the qualitative features of the CEM. The general expression for the
production cross section within the NRQCD factorization approach reads
\beq
d\sigma(H + X) = \sum_n d\hat\sigma(Q\bar Q[n] + X)
\langle{\cal O}^H[n]\rangle
\label{eq-fm}
\eeq Here $d\hat\sigma(Q\bar Q[n] + X)$ describes the short-distance
production of a $Q\bar Q$ pair in the colour/spin/angular momentum
state $n$, and $\langle{\cal O}^H[n]\rangle$, formally a vacuum
expectation value of a NRQCD matrix element (see \cite{bbl} for
details), describes the hadronization of the $Q\bar{Q}$ pair into the
observable quarkonium state $H$. One must note that the cross section
is no more given by a single product of a short-distance and a
long-distance term as in the CSM, but rather by a sum of such terms.
Infrared singularities which eventually show up in some of the
short-distance coefficients would be absorbed into the long-distance
part of other terms, thereby ensuring a well defined overall result.
The relative importance of the various contributions in (\ref{eq-fm})
can be estimated by using NRQCD velocity scaling rules \cite{LMNMH92}.
For $v\to 0$ ($v$ being the average velocity of the heavy quark in the
quarkonium rest frame) each of the NRQCD matrix elements scales with a
definite power of $v$ and the general expression (\ref{eq-fm}) can be
organized into an expansion in powers of $v^2$. For the production of
$S$-wave bound states, like $\jpsi$, the colour-octet matrix elements
associated with higher Fock state components in the quarkonium wave
function are suppressed by a factor of $v^4$ compared to the
colour-singlet contribution. The NRQCD description of $S$-wave
quarkonia production or annihilation thus reduces to the Colour
Singlet Model in the static limit $v= 0$.\footnote{In the case of
  $P$-wave quarkonia, colour-singlet and colour-octet mechanisms
  contribute at the same order in $v$ to annihilation rates and
  production cross sections and must therefore both be included for a
  consistent calculation \cite{BBL92}.}

However, completely ignoring the octet $Q\bar Q$ states, as in the
CSM, can lead to too small quarkonium production cross sections
whenever these states can be copiously produced in the short-distance
interaction in comparison with the colour-singlet ones. This is the
case for large-$p_T$ $\jpsi$ and $\psi'$ production at the Tevatron.
Indeed it has been found \cite{cdfoctet} that gluon production with
subsequent fragmentation into a colour-octet $^3S_1$ $c\bar c$ pair
(which will eventually hadronize into a physical quarkonium) can
successfully describe the experimental data, greatly underestimated by
the CSM. As will be demonstrated below, associated $\jpsi + \gamma$
photoproduction is a particularly sensitive probe of the NRQCD matrix
element $\langle{\cal O}^{\jpsi}[^3S_1,\underline{8}]\rangle$ and will
therefore provide an important consistency check of the colour-octet
explanation of the high-$p_T$ data at the Tevatron.

A recent review of the factorization approach and its applications to
quarkonia production at colliders can be also found in
ref.~\cite{bfc}.

\section{$\jpsi + \gamma$ Production}
\label{psigamma}

Associated $\jpsi + \gamma$ production has been studied in hadronic
collisions by various authors over the recent years \cite{jpg_had}.
Kim and Reya \cite{kr} pointed out that $\jpsi + \gamma$ photo- and
electroproduction represents a powerful tool for discriminating
between the Colour Evaporation Model and the Colour Singlet Model. The
reason for this will be discussed below. Since all previous
calculations were based on the Colour Singlet or Colour Evaporation
Model it is worth to reconsider $\jpsi + \gamma$ production within the
context of the new factorization approach. We will in fact demonstrate
that this reaction is a particularly sensitive probe of the presence
of colour-octet mediated channels in $\jpsi$ production.

\begin{figure}[t]
\begin{center}
\epsfig{file=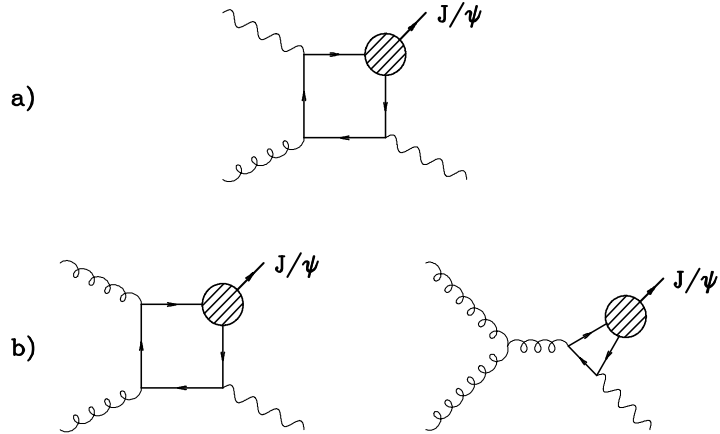,width=9.5cm}
\parbox{12cm}{
\caption{\label{fig1}\small Diagrams contributing within the factorization
  approach to $\jpsi$ + photon associated production in direct (a) and
  resolved (b) photon collisions.}
}
\end{center}
\end{figure}

Before presenting our results within the BBL formalism, let us briefly
review the reason why associated $\jpsi + \gamma$ production provides
a good discrimination between the CEM and the CSM in photoproduction.
Due to the colour constraint of the CSM, the leading order
contribution, fig.~\ref{fig1}a,
\beq
\gamma p \to \gamma + g_p \to c\bar c[^3S_1,\underline{1}] +\gamma 
\to \jpsi + \gamma
\eeq
obviously vanishes in the Colour Singlet Model, since only one gluon 
is attached to the $c\bar c$ line. In order to allow for a second
gluon one must either go to higher order in $\as$ for the direct
photon process given above or consider a {\sl resolved photon
process}:
\beq
\gamma p \to g_\gamma + g_p \to c\bar c[^3S_1,\underline{1}] +\gamma 
\to \jpsi + \gamma
\eeq
In both cases, we can expect a suppression of $\jpsi + \gamma$
photoproduction within the CSM.  This is not the case in the CEM.
Since the colour constraint is absent, the process can proceed via
the direct production channel giving a $c\bar c$ pair of invariant
mass smaller than that of a $D\bar D$ pair threshold:
\beq
\gamma p \to \gamma + g_p \to c\bar c[M_{c\bar c} < M_{D\bar D}] +\gamma 
\to \jpsi + \gamma
\eeq
Kim and Reya have indeed argued that at HERA within the CEM one should
expect about a factor of five more $\jpsi + \gamma$ events than
predicted by the CSM.

As stated before, the factorization approach shares some features of
the CEM by allowing colour-octet intermediate states. For this case,
in particular, it allows the $\jpsi + \gamma$ production to proceed
also via direct photon interactions through a colour-octet $c\bar c$
pair. To be more specific, the following short-distance process can
contribute:

\vspace{.5cm}
\noindent
Direct photon (fig.~\ref{fig1}a):
\beqa
&&\gamma + g_p \to c\bar c[^1S_0,\underline{8}] + \gamma\nonumber \\
&&\gamma + g_p \to c\bar c[^3S_1,\underline{8}] + \gamma\qquad
\hphantom{(\mathrm{Standard~CSM~process})}
\nonumber \\
&&\gamma + g_p \to c\bar c[^3P_J,\underline{8}] + \gamma\nonumber
\eeqa

\noindent
Resolved photon (fig.~\ref{fig1}b):
\beqa\label{partproc}
&&g_\gamma + g_p \to c\bar c[^1S_0,\underline{8}] + \gamma\nonumber \\
&&g_\gamma + g_p \to c\bar c[^3S_1,\underline{1}] + \gamma\qquad
(\mathrm{Standard~CSM~process})\nonumber\\
&&g_\gamma + g_p \to c\bar c[^3S_1,\underline{8}] + \gamma\nonumber \\
&&g_\gamma + g_p \to c\bar c[^3P_J,\underline{8}] + \gamma\nonumber
\eeqa

Light-quark initiated processes are strongly suppressed and can safely
be neglected.  Resolved production of a colour-singlet $\chi_J$ state and a
hard photon, with the $\chi_J$,
radiatively decaying into $\jpsi$ plus an unobserved soft  photon, could
constitute a background to our signal, but turns out to be zero.

According to the NRQCD velocity scaling rules \cite{LMNMH92} and fits
to the Tevatron data \cite{cdfoctet} the colour-octet non-perturbative
matrix elements that enter associated $\jpsi +\gamma$ production
should be suppressed by about two orders of magnitude compared to the
colour-singlet matrix element. Still, however, the two following
features can be qualitatively expected:
\begin{itemize}
\item[i)] the production of colour-octet states via a direct process -
  rather than a resolved one - will at least partially compensate for
  the smaller matrix elements. We should therefore expect a sizable 
  increase in the overall cross section due to colour-octet channels;
\item[ii)] the distribution in the inelasticity of the $\jpsi$, namely
  the ratio of its energy over the initial photon energy in the proton
  rest frame, usually indicated with $z$ (in a generic frame $z =
  p_{\jpsi}\cdot p_p/p_\gamma\cdot p_p$), will be more peaked towards
  one for the colour-octet induced processes. This is again due to the
  presence of a direct photon coupling as opposed to the resolved one,
  where the $g_\gamma$ only carries part of the photon energy into the
  reaction.
\end{itemize}
These qualitative features are in fact born out by our numerical
analysis to be presented and discussed in the next section.  The
parton cross sections (\ref{partproc}) have been evaluated using the
algebraic computer program FORM \cite{form}.  We note here that the
direct photon diagram in fig.~\ref{fig1}a happens to contribute to
$[^3S_1, \underline{8}]$ production only.  Moreover, due to the triple
gluon vertex in the second diagram of fig.~\ref{fig1}b, the cross
sections for the resolved photon contribution to $^1S_0$ and
$^3P_{0,2}$ octet states display a collinear and infrared singularity
in the $p_T=0$ and $z=1$ endpoint.  Rather than properly subtracting
it we will avoid this phase space region by applying a minimum-$p_T$
cut.

\section{Numerical Results}
\label{numres}

The numerical results have been obtained by adopting the following set
of parameters.  As for the strong coupling, we employ the leading
order formula
\beq
\as(\mu) = {{12\pi}\over{(33-2n_f)\ln(\mu^2/\Lambda^2)}}
\eeq
with $n_f = 4$ and $\Lambda_{LO}^{(4)} = 120$ MeV. The scale $\mu$ has
been taken equal to the transverse $\jpsi$ mass, namely $\mu = M_T =
\sqrt{4m_c^2 + p_T^2}$, and the charm mass $m_c$ has been taken equal
to 1.5 GeV. For both the proton and the resolved photon structure
functions the GRV92 leading order parton densities \cite{GRV} have been
employed, evaluated at the factorization scale $\mu_F = M_T$.

\begin{table}
\begin{center}
\begin{tabular}{|c|c|c|} 
\hline
$\langle {\cal{O}}^{J\!/\!\psi}\,[^{3}S_{1},\underline{1}]\rangle 
\hphantom{/m_c^2}$ &$1.16$ GeV$^3$ & $\quad m_c^3 v^3$ \\ 
$\langle {\cal{O}}^{J\!/\!\psi}\,[^{1}S_{0},\underline{8}]\rangle
\hphantom{/m_c^2}$ &$10^{-2}$ GeV$^3$ & $\quad m_c^3 v^7$ \\ 
$\langle {\cal{O}}^{J\!/\!\psi}\,[^{3}S_{1},\underline{8}]\rangle
\hphantom{/m_c^2}$ &$10^{-2}$ GeV$^3$ & $\quad m_c^3 v^7$ \\ 
$\langle {\cal{O}}^{J\!/\!\psi}\,[{}^{3}P_{0},\underline{8}]\rangle / m_c^2$  
& $10^{-2}$ GeV$^3$ & $\quad m_c^3 v^7$ \\
\hline
\end{tabular}
\parbox{12cm}{
\caption{
\label{table1}
\small Values of the NRQCD matrix elements used in the numerical
 analysis, with the velocity and mass scaling. $v$ is the velocity of
 the heavy quark in the quarkonium rest frame. For charmonium it holds
 $v^2 \simeq 0.23$. It also holds 
 $\langle {\cal{O}}^{J\!/\!\psi}\,[{}^{3}P_{J},\underline{8}]\rangle
 \simeq (2J+1)
 \langle {\cal{O}}^{J\!/\!\psi}\,[{}^{3}P_{0},\underline{8}]\rangle$.
}
}
\end{center}
\end{table}

Finally, the non-perturbative matrix elements have been fixed to
values compatible with both the NRQCD scaling rules and with the
Tevatron fits, as in ref.~\cite{ck}. Their values are summarized in
Table~\ref{table1}.

We first compare the total cross sections given by the various
production channels, with a fixed $p_T>1$ GeV cut applied to avoid the
infrared/collinear singularities in some of the cross sections.
Table~\ref{table2} shows the results for a) $\gamma p$ collisions in a
fixed target set-up, with an incoming photon energy of 100 GeV; b)
$\gamma p$ collisions at a center of mass energy of 100 GeV and c)
HERA kinematics, i.e. 27.5 GeV electron and 820 GeV proton collisions.
For the case c) the electroproduction cross sections have been
evaluated by averaging the photoproduction ones weighted by the usual
Weizs\"acker-Williams flux factor,
\beq
\sigma_{ep}(s) = \int_{y_{min}}^{y_{max}} dy f_{\gamma/e}(y) 
\sigma_{\gamma p}(ys)
\eeq
with
\beq
f_{\gamma/e}(y) = {\alpha\over{2\pi}} \left[ {{1 + (1-y)^2}\over{y}}
\ln{{Q_{max}^2}\over{Q_{min}^2}} + 2 m_e^2 y \left( {1\over{Q_{max}^2}}
-{1\over {Q_{min}^2}}\right)\right]
\eeq
where $y=E_\gamma/E_e$, $Q_{min}^2 = m_e^2 y/(1-y)$ and $m_e$ is the 
electron mass. We adopt $Q_{max}^2 = 4$ GeV$^2$  and $y_{min} = 0.15$, 
$y_{max} = 0.86$ according to \cite{zeuswwa}. 

\begin{table}[th]
\begin{center}
\begin{tabular}{|c|c|r|r|r|r|} 
\hline
\multicolumn{2}{|c|}{} & $\gamma p$ & $\gamma p$ & $ep$ & $ep$\\
\multicolumn{2}{|c|}{} & \scriptsize{$E_\gamma$ = 100 GeV} & 
\scriptsize{$\sqrt{s}$ = 100 GeV} &
\scriptsize{$\sqrt{s}$ = 300 GeV} & \scriptsize{$\sqrt{s}$ = 300 GeV}  \\[-4pt]
\multicolumn{2}{|c|}{Channel} & \scriptsize{$p_T>1$ GeV}  
& \scriptsize{$p_T>1$ GeV} & \scriptsize{$p_T>1$ GeV}
& \scriptsize{$p_T>1$ GeV} \\[-4pt]
\multicolumn{2}{|c|}{} &  &  & & \scriptsize{$-3<\eta<3$}\\[-4pt]
\multicolumn{2}{|c|}{} &  &  & & \scriptsize{$E_\gamma > 2$ GeV}\\
\hline
       & $^1S_0,\underline{8}$ &  -- &   -- &  -- & --  \\
Direct & $^3S_1,\underline{8}$ & .48 & 7.67 & .63 & .44 \\
       & $^3P_J,\underline{8}$ &  -- &   -- &  -- & --  \\
\hline
            & $^1S_0,\underline{8}$ & .0013 &   .35 & .044 & .020 \\
            & $^3S_1,\underline{1}$ & .072  & 16.70 & 2.05 & .75  \\
            & $^3S_1,\underline{8}$ & .0012 &   .27 & .033 & .012 \\
Resolved    & $^3P_0,\underline{8}$ & .0046 &  1.03 & .13  & .048 \\
            & $^3P_1,\underline{8}$ & .0005 &   .14 & .018 & .007\\
            & $^3P_2,\underline{8}$ & .0045 &   .97 &  .12 & .043 \\
\hline
\end{tabular}
\parbox{12cm}{
\caption{
\label{table2}
\small Results for the total cross sections (in pb). Notice that while
 in the first data column $E_\gamma$ refers to the incoming photon beam
 energy in the last one it refers instead to the produced outgoing
 photon.} }
\end{center}

\end{table}

The results presented in Table~\ref{table2} clearly show that in a
typical fixed target set-up, with a photon beam energy of 100 GeV, the
cross section is dominated by the direct photon production of a
colour-octet $^3S_1$ state. Although the cross section is small, the
observation of this process at fixed-target experiments would
therefore already provide good evidence for the importance of
colour-octet contributions.

At a higher center of mass energy, on the other hand, the production
of a $\jpsi$ via a colour-singlet $^3S_1$ state in resolved photon
collisions and via a colour-octet $^3S_1$ state in direct photon
collisions represents the major part of the cross section. The numbers
presented in Table~\ref{table2} indicate that at HERA energies
colour-octet channels could amount to about 40\% of the overall
production rate. The presence of colour-octet contributions can
however not be assessed from total cross sections alone, given the
large normalization uncertainties present in the calculation from
higher order corrections, parton distribution functions and charm
quark mass values, as well as unknown higher-twist contributions.

Therefore, we propose to study differential distributions to
disentangle the colour-octet contributions from the standard
colour-singlet one.

\begin{figure}[h,t]
\begin{center}
\hspace*{-1.5cm}
\epsfig{file=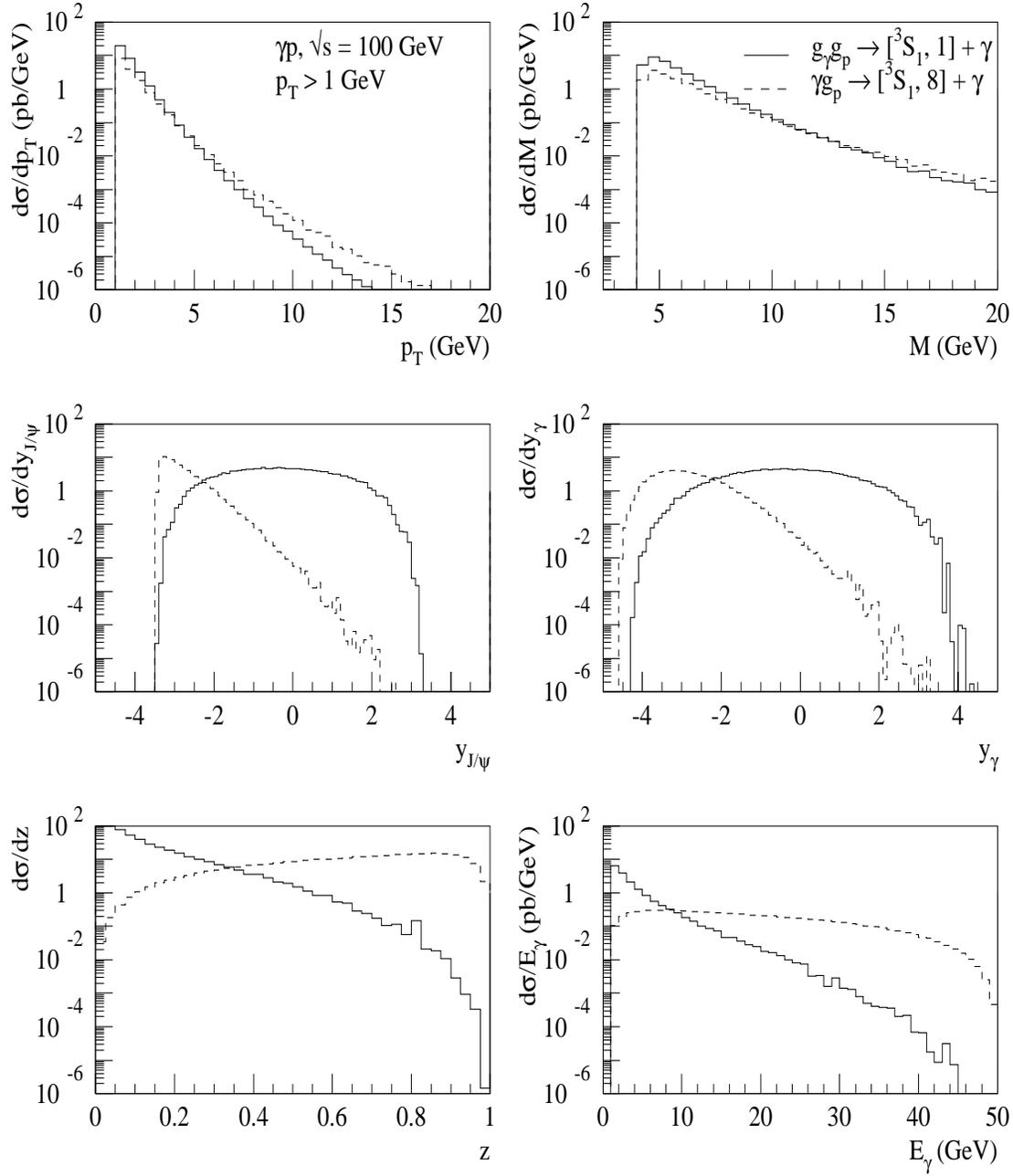,
              bbllx=30pt,bblly=160pt,bburx=540pt,bbury=660pt,
             width=15.cm,height=17.5cm,clip=}
\parbox{12cm}{
\caption{\label{fig2}\small Differential distributions in $\gamma p$ 
  collision at $\protect\sqrt{s} = 100$ GeV. A minimum-$p_T$ cut of 
  1 GeV is applied.  } }
\end{center}
\end{figure}

In fig.~\ref{fig2} we show the differential distributions related to
the total $\gamma p$ cross sections at $\sqrt{s} = $ 100 GeV with a
minimum-$p_T$ cut of 1 GeV, presented in Table~\ref{table2}. The
distributions due to colour-singlet [$^3S_1$,\underline{1}] production
in resolved photon collisions (continuous line) and to colour-octet
[$^3S_1$,\underline{8}] production in direct photon collisions (dashed
line) only are shown. The distributions due to the other colour-octet
processes do indeed present the same features as the ones of
[$^3S_1$,\underline{1}], being also produced in resolved photon
interactions, but are suppressed in magnitude, as can be seen from
Table~\ref{table2}. Their inclusion would therefore not change the
picture we are going to discuss.

As expected, the effect of the colour-octet [$^3S_1$,\underline{8}]
contribution produced in direct photon processes can easily be seen in
at least some of the plots. While the $p_T$ of the $\jpsi$ and the
invariant mass distribution $M$ of the $\jpsi$-$\gamma$ pair are
pretty similar for the colour-singlet and colour-octet induced
channels, the $z$, rapidity and photon energy distributions do indeed
show a strikingly different behaviour.

\clearpage

Recalling that we put ourselves in the so-called ``HERA-frame'', with
the photon (or the electron) traveling in the direction of negative
rapidities, we notice how the direct photon coupling favours the
production of the quarkonium and of the photon in the negative
rapidities region. This contrasts the case of resolved photon
production of colour-singlet $^3S_1$ states, which are uniformly
produced around the central rapidity region.

As for the $z$ distribution, the resolved photon process predicts a
decrease of the cross section going towards the high-$z$ region.. The
direct photon process does on the other hand predict the opposite
behaviour: the cross section now increases going towards $z=1$. The
small dip in the last few bins is due to the minimum-$p_T$ cut.

The photon energy distribution behaves similarly to the $z$
distribution, and is predicted to be much harder in direct photon
processes.

The above distributions (the shapes of which have been checked to be
robust with respect to a higher $p_T$ cut, to be sure of the absence
of $p_T^{min}$ effects) provide a clear experimental signature; in
particular the observation of a substantial fraction of $\jpsi +
\gamma$ events in the high-$z$ region would already be good evidence
for the presence of colour-octet contributions to the $\jpsi$
production cross section.

\begin{figure}[h,t]
\begin{center}
\hspace*{-1.5cm}
\epsfig{file=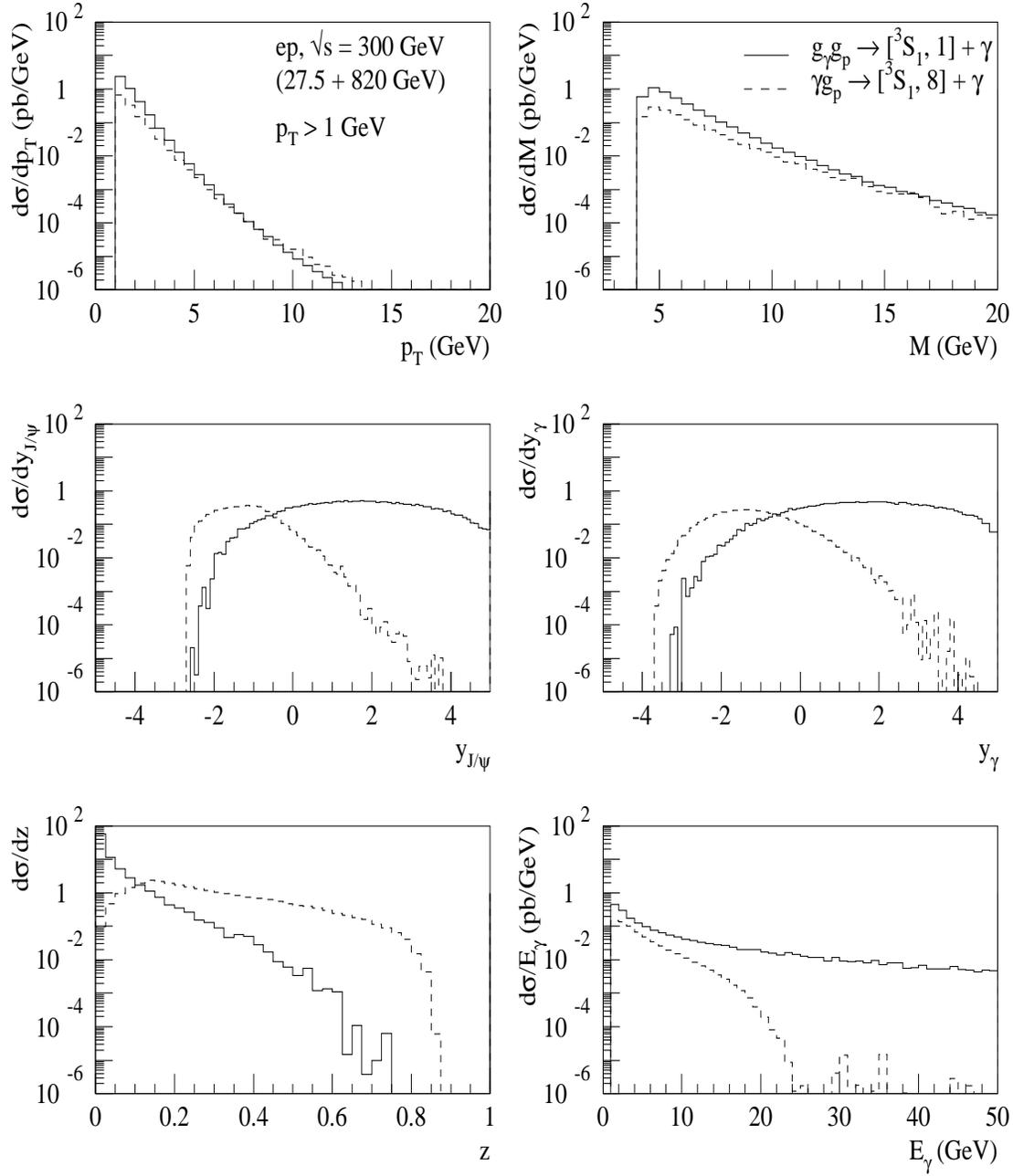,
              bbllx=30pt,bblly=160pt,bburx=540pt,bbury=660pt,
            width=15.cm,height=17.5cm,clip=}
\parbox{12cm}{
\caption{\label{fig3}\small Differential distributions in $ep$ collision
  at $\protect\sqrt{s} = 300$ GeV. A minimum-$p_T$ cut of 1 GeV is
  applied.  } }
\end{center}
\end{figure}

However, since actually HERA provides electron-proton interactions, we
are going to investigate how the distributions look like in this
frame. For consistency, we will first consider, in fig.~\ref{fig3},
the differential cross section corresponding to the total $ep$ cross
section shown in Table~\ref{table2}.  Successively, the effect of some
experimental-like cuts will be taken into account.

Comparing fig.~\ref{fig3} with fig.~\ref{fig2} we notice that, due to
the system being now boosted in the proton direction (i.e.\ positive
rapidities), the non Lorentz invariant observables are affected. More
precisely, the rapidity distributions of both the $\jpsi$ and of the
photon are slightly smeared and shifted towards positive rapidities.
The largest difference can however be observed in the photon energy
distributions: the one due to the resolved photon process is hardened
with respect to $\gamma p$ interactions, while the one due to direct
photon interaction is greatly softened. This can be understood
assuming that most of the very energetic photons in direct photon
collisions are produced in the incoming photon direction (see the
rapidity distribution). The effect of a boost of the event in the
opposite direction will then soften this distribution. In resolved
photon interactions, on the other hand, events of this kind are
produced more uniformly around the central rapidity region, and the
boost will then also harden at least part of them.

\begin{figure}[h,t]
\begin{center}
\hspace*{-1.5cm}
\epsfig{file=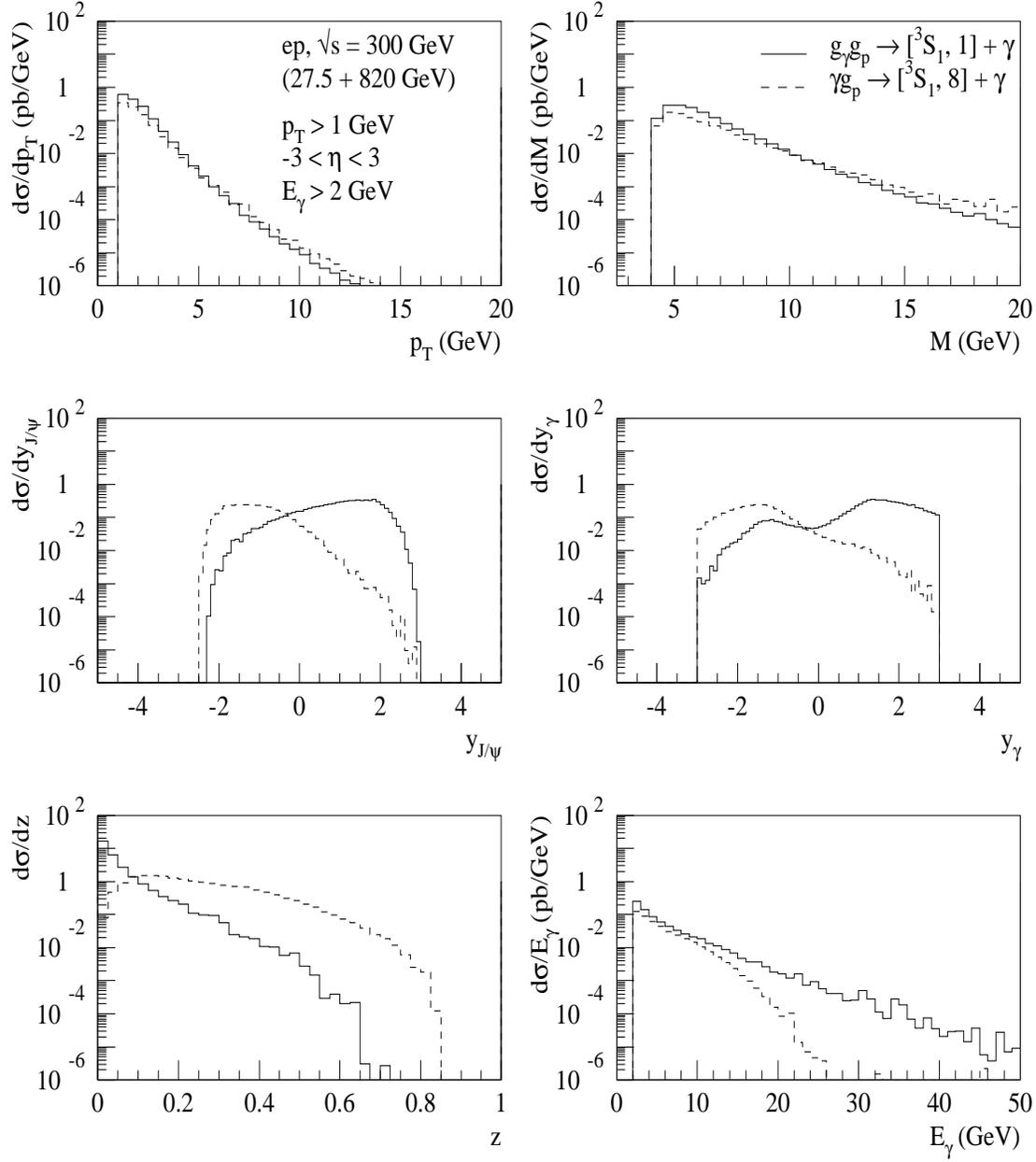,
              bbllx=30pt,bblly=160pt,bburx=540pt,bbury=660pt,
            width=15.cm,height=16.9cm,clip=}
\parbox{12cm}{
\caption{\label{fig4}\small Differential distributions in $ep$ collision
  at $\protect\sqrt{s} = 300$ GeV. A minimum-$p_T$ cut of 1 GeV, a
  minimum outgoing photon energy cut of 2 GeV and a pseudorapidity cut
  $|\eta_{\jpsi,\gamma}| < 3$ are applied.  } }
\end{center}
\end{figure}

\clearpage

The effect of applying some realistic experimental cuts can be
appreciated in fig.~\ref{fig4}. Since the produced photon must be
clearly visible in the event, it is now required to lie (together with
the $\jpsi$) within the pseudorapidity region $|\eta|<3$ and to have
an energy, in the lab frame, greater than 2 GeV.  This last
requirement also ensures that it will not be mistaken with a photon
coming from the radiative decay of a $P$-wave state to a $\jpsi$,
having energies of a few hundreds MeV. A further experimental
selection criterion to exclude radiative decays photons consists in
asking the photon $p_T$ to be roughly opposite to that of the $\jpsi$.
The distributions of fig.~\ref{fig4} are qualitatively similar to the
ones - without cuts - of fig.~\ref{fig3}, showing that colour-octet
contributions in associated $\jpsi + \gamma$ electroproduction should
be visible in the photon fragmentation region (negative rapidities in
the HERA frame) or, more clearly, in the large-$z$ region.

\section{$\jpsi + \gamma$ in hadron collisions}

Before closing, it is worth noticing that the resolved processes of
fig.~\ref{fig1}b also contribute to the associated $\jpsi + \gamma$
production in hadron collisions. To check whether a signature for
colour-octet mediated processes can be seen at a hadron collider we
have evaluated the total $\jpsi + \gamma$ cross section in $p\bar p$
collisions at $\sqrt{s} = 1800$ GeV. Light-quark initiated processes
are suppressed and have been neglected. The following cuts have been
imposed:
\beq p_T > 4~\mbox{GeV}, \qquad |\eta_{\jpsi,\gamma}| < 0.6,
\qquad E_\gamma > 2~\mbox{GeV} \eeq 
The results are displayed in Table~\ref{table3}. The colour-octet
terms can be seen to enhance the cross section by about 50\%. However,
given the large normalization uncertainties involved and the fact that
the differential distributions in this case do not differ
substantially from the colour-singlet process' ones, we conclude that
there is little hope to shed light on the colour-octet mechanism via
$\jpsi + \gamma$ production in hadron collisions.

\begin{table}[th]
\begin{center}
\begin{tabular}{|c|c|r|} 
\hline
{}        & $\sigma_{p\bar p}(\jpsi+\gamma)$ (pb) \\
{Channel} & $\sqrt{s}$ = 1800 GeV  \\
{}        & {$p_T>4$ GeV}   \\
{}        & {$|\eta_{\jpsi,\gamma}| < 0.6$}   \\
{}        & {$E_\gamma > 2$ GeV}   \\
\hline
             $^3S_1,\underline{1}$ & 93.2    \\
             $^1S_0,\underline{8}$ & 12.3  \\
             $^3S_1,\underline{8}$ & 1.5  \\
             $^3P_0,\underline{8}$ & 10.0  \\
             $^3P_1,\underline{8}$ & 13.8 \\
             $^3P_2,\underline{8}$ & 7.7   \\
\hline
\end{tabular}
\parbox{12cm}{
\caption{
\label{table3}
\small Results for the total cross sections in hadron collisions (in
       pb).} }
\end{center}
\end{table}

\section{Conclusions}
\label{concl}

The associated $\jpsi + \gamma$ production in $ep$ and $\gamma p$
collisions has been proposed as a powerful tool for establishing the
presence of quarkonia production processes mediated by colour-octet
$Q\bar Q$ states, as suggested by the factorization approach of
Bodwin, Braaten and Lepage.

The very fact that colour-octet contributions to $\jpsi + \gamma$
photoproduction can proceed also via a direct photon coupling rather
than via a resolved one only - as for the colour-singlet channel -
leads to clean experimental signatures.  We have shown that at HERA
energies the colour-octet terms can increase the cross section by
about 50\% and, most importantly, produce a $\jpsi$ energy
distribution $d\sigma/dz$ strikingly different from the one predicted
by the colour-singlet channel alone.

The hadroproduction case, taking the Tevatron as an example, has also
been investigated. This cross section is also increased by about 50\%
by octet terms, but no significant signatures in differential
distributions can be found.

\vspace*{5mm}

{\it Note added:} After completion of our work an analysis of
colour-octet contributions to associated $\jpsi +\gamma$ production in
hadron-hadron collisions appeared \cite{clsjpg}. After correcting what
is probably a trivial misprint in the labelling of figs.~6-8 of
\cite{clsjpg} (they should have units nb rather than pb), their
results are consistent with ours.

\newcommand{\zp}[3]{Z.\ Phys.\ {\bf C#1} (19#2) #3}
\newcommand{\pl}[3]{Phys.\ Lett.\ {\bf B#1} (19#2) #3}
\newcommand{\plold}[3]{Phys.\ Lett.\ {\bf #1B} (19#2) #3}
\newcommand{\np}[3]{Nucl.\ Phys.\ {\bf B#1} (19#2) #3}
\newcommand{\prd}[3]{Phys.\ Rev.\ {\bf D#1} (19#2) #3}
\newcommand{\prl}[3]{Phys.\ Rev.\ Lett.\ {\bf #1} (19#2) #3}
\newcommand{\prep}[3]{Phys.\ Rep.\ {\bf C#1} (19#2) #3}
\newcommand{\niam}[3]{Nucl.\ Instr.\ and Meth.\ {\bf #1} (19#2) #3}
\newcommand{\mpl}[3]{Mod.\ Phys.\ Lett.\ {\bf A#1} (19#2) #3}


\begin{thebibliography}{999}

\bibitem{bbl} G.T.~Bodwin, E.~Braaten and G.P.~Lepage, \prd{51}{95}{1125}

\bibitem{cdf} F.~Abe {\it et al.} (CDF Coll.), \prl{69}{92}{3704},
           \prl{71}{93}{2537}, FERMILAB-CONF-96/156-E

\bibitem{cdfoctet} E.~Braaten and S.~Fleming, \prl{74}{95}{3327}\\
               M.~Cacciari, M.~Greco, M.L.~Mangano and A.~Petrelli, 
               \pl{356}{95}{560}\\
               P.~Cho and A.K.~Leibovich, \prd{53}{96}{150} and 
               \prd{53}{96}{6203}\\
               M.~Cacciari and M.~Greco, \prl{73}{94}{1586}\\
               E.~Braaten, M.A.~Doncheski, S.~Fleming and M.L.~Mangano, 
                 \pl{333}{94}{548}\\
               D.P.~Roy and K.~Sridhar, \pl{339}{94}{141}
               
\bibitem{ee} E.~Braaten and Y.-Q.~Chen, \prl{76}{96}{730}

\bibitem{zdecay} K.~Cheung, W.-Y.~Keung and T.C.~Yuan, \prl{76}{96}{877}\\
                 P.~Cho, \pl{368}{96}{171}

\bibitem{coft} W.-K.~Tang and M.~V\"{a}nttinen, \prd{53}{96}{4851} and
               \prd{54}{96}{4349}\\
               S.~Fleming and I.~Maksymyk, \prd{54}{96}{3608}\\
               S.~Gupta and K.~Sridhar, \prd{54}{96}{5545}\\
               M.~Beneke and I.Z.~Rothstein, \prd{54}{96}{2005}

\bibitem{bdecays} P.~Ko, J.~Lee, and H.S.~Song, \prd{53}{96}{1409}\\
               S.~Fleming, O.F.~Hern\'{a}ndez, I.~Maksymyk and H.~Nadeau,
               MADPH-96-953 (hep-ph/9608413)

\bibitem{ck} M.~Cacciari and M.~Kr\"amer, \prl{76}{96}{4128}

\bibitem{afm} J.~Amundson, S.~Fleming and I.~Maksymyk, UTTG-10-95
              (hep-ph/9601298)

\bibitem{kls} P.~Ko, J.~Lee and H.S.~Song, \prd{54}{96}{4312}

\bibitem{gks} R.~Godbole, D.P.~Roy, and K.~Sridhar, \pl{373}{96}{328}

\bibitem{ma} J.P.~Ma, \np{460}{96}{109}

\bibitem{ckhera} M.~Cacciari and M.~Kr\"amer, Proceedings
                 of the workshop {\it Future Physics at HERA}, Vol 1, p. 416,  
                 eds.\ G.~Ingelman, A.~De~Roeck and R.~Klanner 
                 (hep-ph/9609500)

\bibitem{h1} S.~Aid {\it et al.} (H1 Coll.), \np{472}{96}{3}. See also 
             M.~Derrick {\it et al.} (ZEUS Coll.), presented by L.~Stanco
             at the International Workshop on Deep Inelastic Scattering,
             Rome, April 1996

\bibitem{CSS} J.C.~Collins, D.E.~Soper and G.~Sterman, \np{263}{86}{37}

\bibitem{CEM} H.~Fritzsch, \plold{67}{77}{217}\\
              F.~Halzen, \plold{69}{77}{105}\\
              F.~Halzen and S.~Matsuda, \prd{17}{78}{1344}\\
              M.~Gl\"uck, J.~Owens and E.~Reya, \prd{17}{78}{2324}

\bibitem{aeg} A.~Bramon, E.~Etim and M.~Greco, \plold{41}{72}{609}

\bibitem{CSM} E.L.~Berger and D.~Jones, \prd{23}{81}{1521}\\
              R.~Baier and R.~R\"{u}ckl, \plold{102}{81}{364}\\
              for a recent review see also
              G.A.~Schuler, CERN-TH.7170/94 (hep-ph/9403387) 

\bibitem{mlm} M.L.~Mangano, in {\sl Tenth Topical Workshop on
 Proton-Antiproton Collider Physics}, eds. R. Raja and J. Yo (American
 Institute for Physics, 1995), CERN-TH-95-190 (hep-ph/9507353)

\bibitem{LMNMH92} G.P.~Lepage, L.~Magnea, C.~Nakhleh, U.~Magnea and
                  K.~Hornbostel, \prd{46}{92}{4052}

\bibitem{BBL92} G.T.~Bodwin, E.~Braaten and G.P.~Lepage, 
                \prd{46}{92}{R1914}

\bibitem{bfc} E.~Braaten, S.~Fleming and T.C.~Yuan, OHSTPY-HEP-T-96-001 
              (hep-ph/9602374)
               
\bibitem{jpg_had} M.~Drees and C.S.~Kim, \zp{53}{92}{673}\\
            K.~Sridhar, \pl{289}{92}{435}\\
            R.V.~Gavai, R.M.~Godbole and K.~Sridhar, \pl{299}{93}{157}\\ 
            K.~Sridhar, \prl{70}{93}{1747}\\
            E.L.~Berger and K.~Sridhar, \pl{317}{93}{443}\\
            M.A.~Doncheski and C.S.~Kim, \prd{49}{94}{4463}\\
            C.S.~Kim and E.~Mirkes, \pl{346}{94}{124} and 
            \prd{51}{95}{3340}\\
            D.P.~Roy and K.~Sridhar, \pl{341}{95}{413}\\
            H.A.~Peng, Z.M.~He and C.S.~Ju, \pl{351}{95}{349}
   
\bibitem{kr} C.S.~Kim and E.~Reya, \pl{300}{93}{298}

\bibitem{form} FORM~2.0 by J.A.M.~Vermaseren, CAN, Amsterdam, 1991

\bibitem{GRV} M.~Gl\"{u}ck, E.~Reya, and A.~Vogt, \prd{46}{92}{1973}, 
 \zp{53}{92}{127} and \zp{67}{95}{433} 

\bibitem{zeuswwa} M.~Derrick {\it et al.} (ZEUS Coll.), \pl{349}{95}{225}

\bibitem{clsjpg} C.S.~Kim, J.~Lee, and H.S.~Song, KEK-TH-474 
                (hep-ph/9610294)

\end{thebibliography}
\end{document}